\documentclass{webofc}
\usepackage[varg]{txfonts}   
\newcommand{\Planck}{{\it Planck}}

\begin{document}
\title{CMB at small scales: Cosmology from tSZ power spectrum}
%
%

\author{\firstname{Marian} \lastname{Douspis}\inst{1}\fnsep\thanks{\email{marian.douspis@ias.u-psud.fr}} \and 
	\firstname{Laura} \lastname{Salvati}\inst{1} \and
	\firstname{Adélie} \lastname{Gorce}\inst{1,2}
	\and
	\firstname{Nabila} \lastname{Aghanim}\inst{1}
}

\institute{Institut d'Astrophysique Spatiale, CNRS, Université Paris Saclay, Orsay, France 
\and
          Department of Physics and McGill Space Institute, McGill University, Montreal, QC, Canada H3A 2T8           }

\abstract{%
Small scale CMB angular power spectrum  contains not only primordial CMB information but also many contaminants coming from secondary anisotropies. Most of the latter depend on the cosmological model but are often marginalised over. We propose a new analysis of the SPT data focusing on the cosmological contribution of the thermal Sunyaev Zel'dovich (tSZ) effect. We model the tSZ angular spectrum with the halo model and train a random forest algorithm to speed up its computation. We show that using the cosmological information of the tSZ on top of the primordial CMB one contained in SPT data bring more constraints on cosmological parameters. We also combine for the first time Planck tSZ angular power spectrum with SPT ones to put further constraints. This proof of concept study shows how much a proper modelling of the foregrounds in the cosmological analyses is needed.
}
\maketitle
\section{Introduction}
\label{intro}
The thermal SZ effect \cite{sunyaev_zeldovich_1980} is a good tracer of the hot gas in the Universe, and is often use to detect or characterise clusters of galaxy.  It is also a secondary anisotropy which contaminates CMB data and analyses. Frequency maps observed in the millimetre range contain the sum of many signals coming form the last scattering surface (primordial CMB) and from all gravitational and electromagnetic effects along the line of sight (one of which is the tSZ). Using an adapted component separation method, it is possible to retrieve the tSZ effect from several frequency map and so create a map tracing the hot gas \cite{Planck13,Planck2015_tsz_map}. For example, using for the first time data from two different experiments, Planck and ACT, \cite{PACT} were able to reconstruct an optimal tSZ map probing both large scales (from Planck) and small scales (from ACT).  From the Planck tSZ map, the angular power spectrum can be estimated and further used to put constraints on the cosmological parameters (eg. \cite{Salvati2018}). 

If no component separation is applied, each frequency map and thus each corresponding angular power spectrum, contains a contribution from tSZ. It is consequently necessary to be able to model all the contributions (including the tSZ angular power spectrum) to perform cosmological analyses. 
Most current analyses of CMB observations at small scales are done following the same approach  \citep[see, e.g.,][]{George:2014oba, 2017A&A...597A.126C, Reichardt2020, 2021JCAP...01..031H}. A theoretical CMB power spectrum is added to templates of each non-CMB signal to reproduce the observed power spectrum. As mentioned above, apart from the tSZ, the remaining non-CMB components are kinetic SZ, thermal dust emission from dusty star-forming galaxies (DSFG) -- both the Poisson and spatially clustered component, the radio galaxy emission, the Galactic cirrus signal and the cosmic infrared background (CIB). The cosmological parameters enter only the CMB power spectrum, whilst the amplitudes of all templates are set free and marginalised over:
\begin{equation}\label{eq:clsobs}
C_{\ell}^{\text{obs}} \equiv C_{\ell}^{\text{CMB}} (\Theta) + \sum_i D_{3000,i} \times T_i,
\end{equation}
where $\Theta$ is the set of free cosmological parameters and $D_{3000,i}$ is the free amplitude of a template $T_i$ for foreground component $i$, normalised to one at $\ell=3000$. 
All but CMB signals are con- sidered nuisance quantities and the cosmological information is retrieved only from the CMB part. There are two disadvantages to this approach. First, no cosmological information is retrieved from the non-CMB contribution while tSZ, kSZ, CIB contain such information. Second, the analysis is most often not coherent: the template of contribution i is obtained using a given cosmological parameter set which is different from the one used for computing the template j. 

Our goal is thus to propose an approach which alleviate these two imperfections, by doing a coherent analysis retrieving the maximum of the cosmological information from all contributions (primordial and secondary). In this first study, we focus on using the tSZ as an example of foreground containing cosmological information, and leave the study of kSZ signal \cite{PAPER2} or a complete analysis of the combination of Planck and SPT \cite{PAPER3} to companion papers. Our approach consists in substituting templates of non CMB signals with models depending on cosmological parameters.  We will thus assume:
\begin{equation}
C_{\ell}^{\text{obs}} \equiv C_{\ell}^{\text{CMB}} (\Theta) + C_{\ell}^{\text{tSZ}}(\Theta, \Sigma)+ \sum_j D_{3000,j} \times T_j
\end{equation}
where $\Sigma$ is the set of scaling relation parameters and $j$ runs over all the remaining non-CMB components. 

We apply our approach to the SPT power spectra from \cite{Reichardt2020} where the tSZ  dominates the CMB signal. A full analysis is presented in \cite{Douspis21} but we will focus on the approximated modelling of the tSZ angular spectrum in the next section, the data we use in section \ref{sec-3} and the main results in section \ref{sec-4}.

\section{tSZ power spectrum}
\label{sec-2}

The derivation of the tSZ power spectrum based on the halo model is described in \cite{Douspis21, Salvati2018} and citations therein. It implies the computation and integration of the Fourier transform of the Compton-y parameter over a large range of masses, redshifts and k-modes. As the multipole range increases towards higher $\ell$ with high resolution instruments like SPT, the time to compute the $C_\ell^{tSZ}$’s increases and slows down the sampling. 

Therefore, we turn to machine learning techniques to approximate the full "true" $C_\ell$ ’s computed with the halo model. We choose Random Forest (RF) algorithm for its simplicity, rapidity and ability to learn non linear relation between inputs and outputs \cite[e.g.][]{Bonjean2019, RFex1,RFex2}. We train the RF  with 15000 models for which we give as input 5 cosmological parameters and 3 scaling relation parameters, and as output the values of the $C_\ell^{tSZ}$’s computed with the halo model at 10 given $\ell$'s. Then, we interpolate these 10 values to obtain the atSZ angular power spectrum at all $\ell$'s   needed by the SPT likelihood. Figure~1 shows the comparison of 10 predicted $C_\ell$'s versus the test ones. The score is 96.2\% and the precision lower than 2\%, low enough for today's observations.

\begin{figure}[h]
\begin{center}
\includegraphics[scale=0.3]{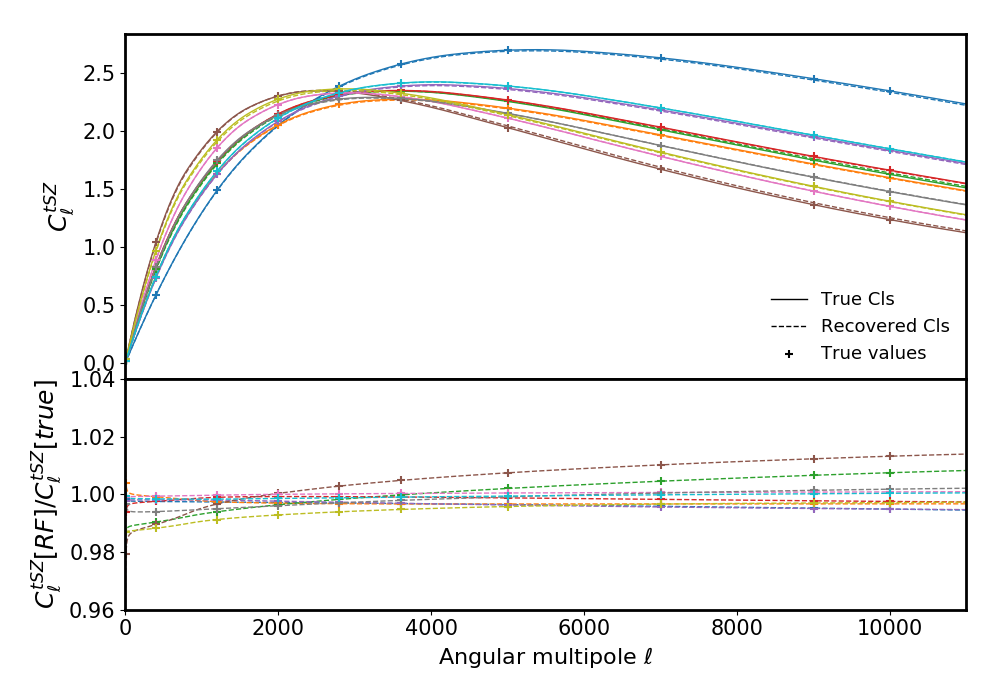}
\label{fig-RF}       
\caption{Comparison of the `true' tSZ $C_\ell$'s values at given $\ell$ (crosses, and solid lines) with the RF-inferred $C_\ell$'s after interpolation (dashed lines) for a sample of models (units are arbitrary before renormalisation in $y$-units at 143~GHz). The lower part of the plot shows the ratio of RF $C_\ell$'s over `true' $C_\ell$'s. }
\end{center}
\end{figure}

 \section{Data and method}
\label{sec-3}

We consider the observed signal on small scales (high multipoles $\ell$) and, therefore, make use of the SPT temperature-only data and likelihood introduced in \cite{Reichardt2020} and made publicly available by the authors. We refer the interested reader to this work for details on the dataset, but highlight the fact that the data considered spans the range $2000 \le \ell \le 11000$ at frequencies 95, 150 and 220 GHz. The likelihood makes use of auto- and cross-spectra and marginalises over calibration and beam parameters. In the baseline analysis of Reichardt et al., the total signal is modelled by CMB, tSZ, kSZ, galactic cirrus contamination, radio and infrared galaxies spectra and tSZ-CIB cross-spectra. In this work, we modified only the tSZ spectra and consequently the tSZ-CIB cross spectra contributions, computed in the SPT likelihood as the sampled correlation co- efficient times a function of the tSZ Cl’s (as defined in Zahn et al. 2012). Implementing new CIB and kSZ modelling will be the focus of upcoming works. 

We have modified and used the publicly available Monte Carlo Markov Chain CosmoMC code \cite{cosmomc} in order to compute the tSZ spectrum with our RF implementation. We consider five cosmological parameters : the baryon and cold dark matter densities, $\Omega_b h^2$ and $\Omega_c h^2$; the ratio of the sound horizon to the angular diameter distance at decoupling $\theta$; the scalar spectral index, $n_s$; the overall normalisation of the spectrum, $A_s$;  the reionisation optical depth $\tau$ is assumed fixed. For the scaling relations, we sample three parameters :the mass bias $(1-b)$, the mass slope $\alpha$ and the normalisation $Y_*$. We assume the self-similar evolution for halos, and fix the redshift evolution to  $2/3$.
We focus here our results on the  parameters most sensitive to the tSZ effect: $\Omega_m$, $\sigma_8$, $H_0$ , and the mass bias parameter $(1-b)$ while others are marginalised over. We assume Gaussian priors on $n_s$ and  $\Omega_b h^2$  as given by Planck.

\section{Results}
\label{sec-4}

We first apply our approach on SPT data alone and study the consequences of substituting the tSZ template with cosmology-dependent spectra. Figure~2 compares the constraints  in the two configurations. We can see that using SPT data alone with the tSZ template (in black) allows for a large range of parameter values, including a narrow degeneracy between $H_0$ and $\Omega_m$. These constraints are driven by the CMB spectrum dominating other signals on the range $2000 < \ell < 4000$ but being negligible at smaller scales. As expected, substituting the template by our RF tSZ power spectrum allows to exploit the full cosmological information (in green), and the constraints are largely tightened with a factor roughly two of improvement.

\begin{figure}[h]
\centering
\sidecaption
\includegraphics[scale=0.3]{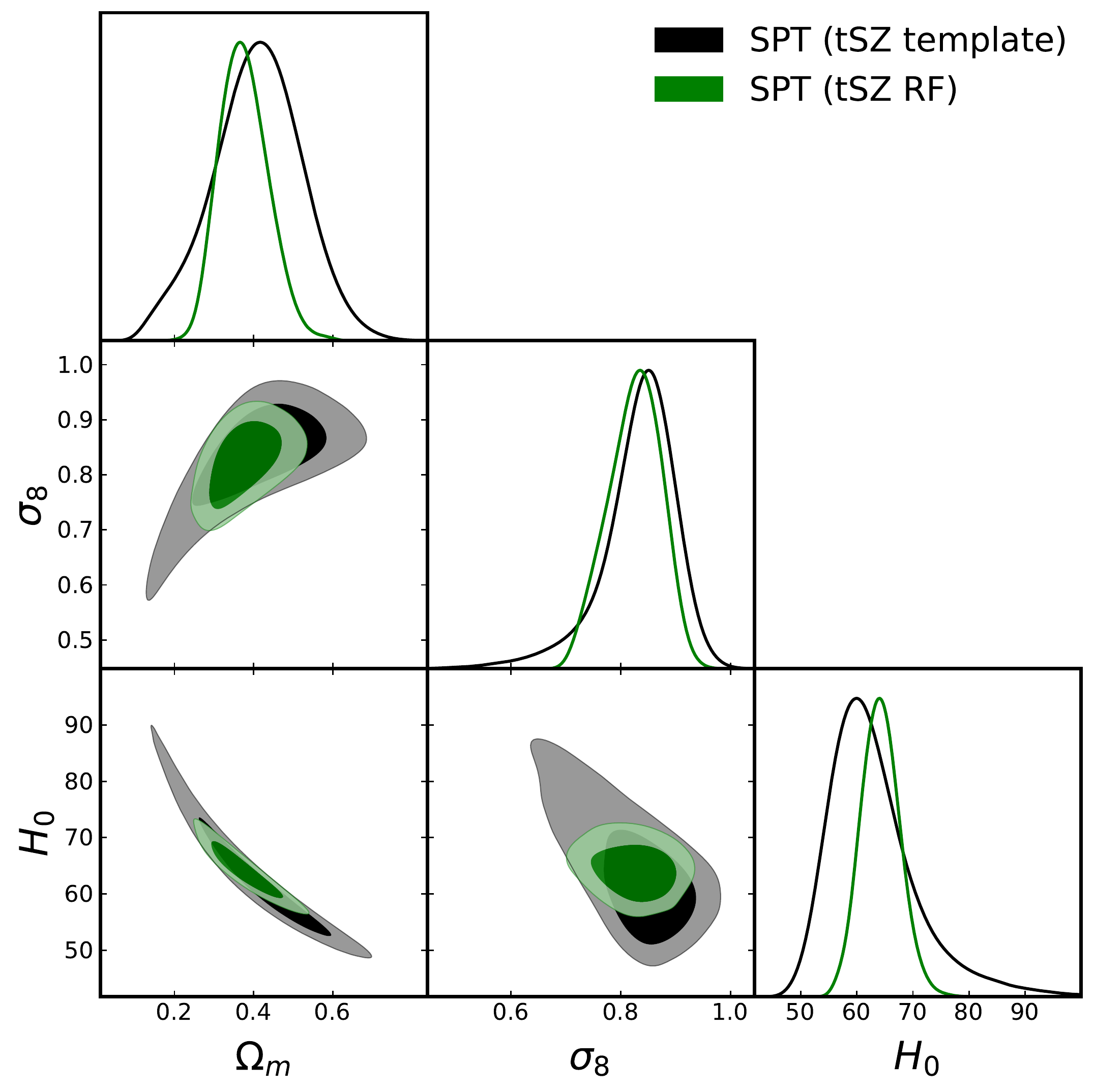}
\label{fig-CP}       
\caption{Constraints on cosmological parameters obtained with SPT high-$\ell$ data using the tSZ template as done in \cite[][in black]{Reichardt2020} and using the tSZ power spectra derived from the halo model (in green). }
\end{figure}

We then add the Planck tSZ angular power spectrum \cite{Planck2015_tsz_map} in the likelihood to add contraints coming from large scale ($100 < \ell < 1000$). We only consider the case using the RF modelling of the tSZ spectrum. Adding Planck tSZ data does not improve drastically the constraints (Fig.~3), but move the best parameters within 1-$\sigma$. $\Omega_m$ and  $\sigma_8$ are shifted to lower values, while the mass bias parameter is shifted high, towards a better agreement with the combination of Planck CMB and cluster number counts \cite{Salvati2018}. The constraints are still not good enough anyway to determine the mass bias. We thus study how the constraints vary when adding a Gaussian prior on the latter (from CCCP \cite{Hoekstra2015}). The degeneracies are broken, and low values of  $\Omega_m$ and  $\sigma_8$  are preferred. All results are summarise in Table~\ref{tab-1}.

\begin{figure}[h]
\centering
\sidecaption
\includegraphics[scale=0.3]{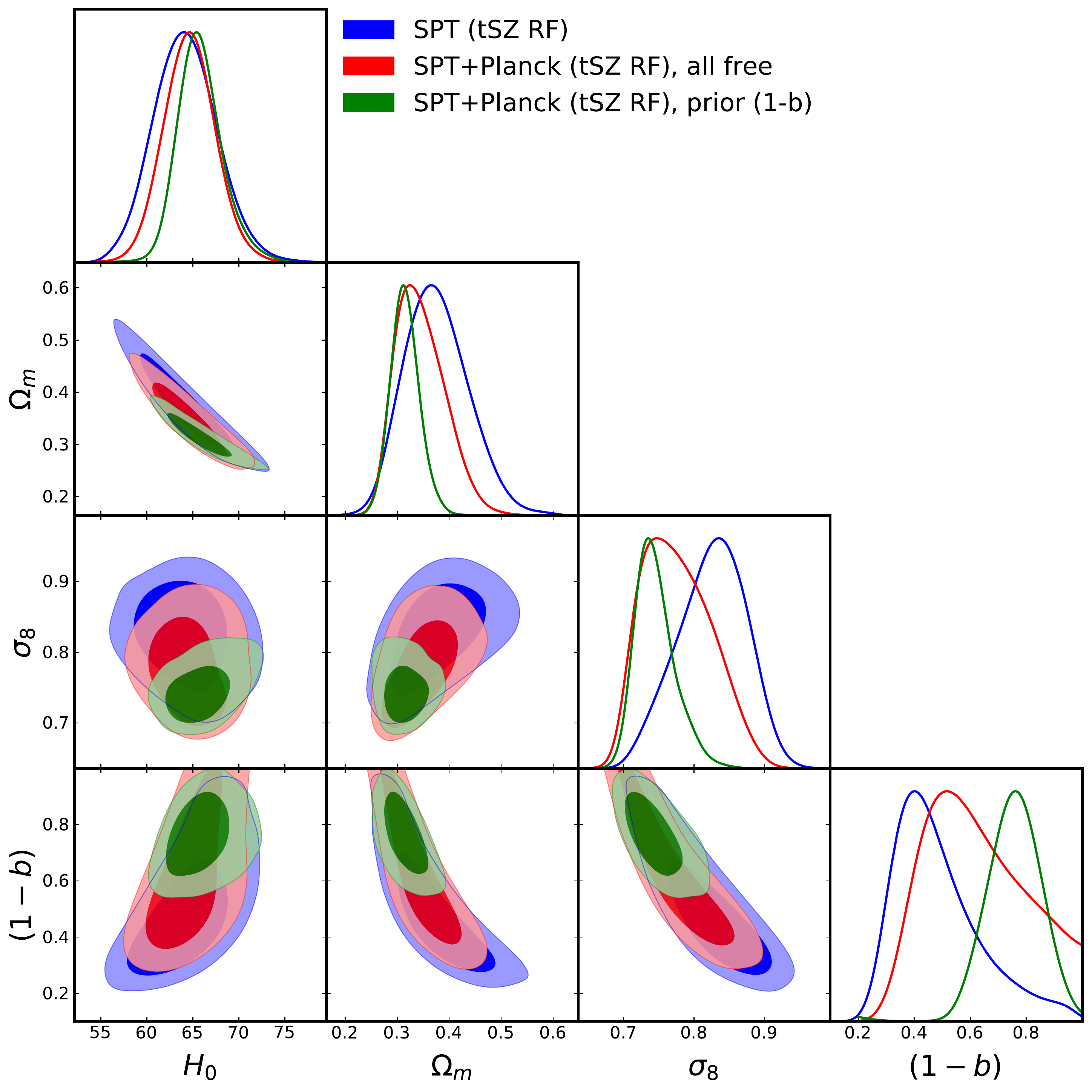}
\label{fig-2CP}       
\caption{Constraints on cosmological and scaling relation parameters using different data sets and priors: in blue, SPT high-$\ell$ data alone, and in red, with \Planck~tSZ power spectra added. Results when CCCP prior on the mass bias is added are shown in green. }
\end{figure}

\begin{table*}
    \centering
\begin{tabular}{c|c|c|c|c}
\tiny
 & SPT (tSZ template) & SPT (tSZ RF) & SPT + Planck (RF) [U] & SPT + Planck (RF) [G] \\ 
 \hline
 $\Omega_m$ & $0.40 \pm 0.11$ & $ 0.38 \pm 0.06$ & $ 0.35 \pm 0.05$ & $ 0.31 \pm 0.03$ \\
 \hline
 $\sigma_8$ & $0.83 \pm 0.07$ & $ 0.825 \pm 0.048$ & $0.774 \pm 0.048 $& $0.744 \pm 0.027$\\
 \hline
 $H_0$ & $63.2 \pm 8.12$ & $64.2 \pm 3.4$ & $ 64.4 \pm 2.8$ & $65.9 \pm 2.35 $\\
 \hline
 $(1-b)$ & \textbf{-} & $ 0.49 \pm 0.16$ & $ 0.62 \pm 0.17$ & $0.76 \pm 0.09$\\
 \hline
\end{tabular}
    \caption{Constraints on cosmological and mass bias parameters for the different dataset combinations. We report the $68\%$ confidence levels. [U] goes for uniform prior on (1-b) while [G] goes for the CCCP Gaussian prior, $(1-b)=0.79\pm0.09$, applied on (1-b)}
    \label{tab-1}
\end{table*}

\section{Conclusions}
\label{sec-5}

This work is the first attempt to bring all full cosmological information from secondary anisotropies and foregrounds in CMB analysis at small scales. In this study we focused on tSZ effect only and SPT data. We have shown that substituting the tSZ template by a cosmology dependent spectrum brings more constraints on cosmological parameters. Adding the Planck measurement of tSZ spectrum at large scales does not improve much the constraints but eliminates too large or too small values of the parameters (eg. $\Omega_m=0.5$). Such a combination prefers low values of $\sigma_8~0.77$ compared to \Planck\ CMB analysis. We are now proceeding in implementing kSZ and CIB  cosmology dependent spectra in the likelihood (Gorce et al, Douspis et al., in prep.). On the methodological aspect, we built a modelling of the tSZ spectrum using machine learning technique (Random Forest) and implemented it in the COSMOMC framework. We gained a factor  more than 100 in the computation of the spectra and a factor of 10 in running MCMC chains. 
Such approach (fast and coherent), maybe  of great interest in the futur not only for constraining cosmological parameters but also astrophysical ones describing the properties of the large scale structures.

%

\bibliography{biblio}

%
%
%
%

\end{document}